\documentclass[10pt,letterpaper]{article}
\usepackage[top=0.85in,left=2.75in,footskip=0.75in]{geometry}
\usepackage{amsmath,amssymb}
\usepackage{changepage}
\usepackage{textcomp,marvosym}
\usepackage{cite}
\usepackage{nameref,hyperref}
\usepackage[right]{lineno}
\usepackage[nopatch=eqnum]{microtype}
\DisableLigatures[f]{encoding = *, family = * }
\usepackage[table]{xcolor}
\usepackage{array}
\newcolumntype{+}{!{\vrule width 2pt}}
\newlength\savedwidth

\raggedright
\usepackage[aboveskip=1pt,labelfont=bf,labelsep=period,justification=raggedright,singlelinecheck=off]{caption}

\makeatletter
\renewcommand{\@biblabel}[1]{\quad#1.}
\makeatother
\usepackage{lastpage,fancyhdr,graphicx}
\usepackage{epstopdf}
\fancyheadoffset[L]{2.25in}
\fancyfootoffset[L]{2.25in}
\begin{document}
\vspace*{0.2in}
\begin{flushleft}
{\Large
\textbf\newline{Impact of Network Size and Frequency of Information Receipt on Polarization in Social Networks}
}
\newline
\\
Sudhakar Krishnarao\textsuperscript{1\Yinyang}*,
Shaja Arul Selvamani\textsuperscript{2\Yinyang}*
\\
\bigskip
\textbf{1} Retired, Wayanad, Kerala, India.
\\
\textbf{2} Outshift by Cisco Systems Inc, USA.
\\
\bigskip
\Yinyang These authors contributed equally to this work.
* sudhakar.iitb@gmail.com,
* sharulse@cisco.com
\end{flushleft}

\section*{Preprint Notice}
This is the preprint version of an article published in Complexity journal. For citation purposes, please refer to the final published version of the article:
https://doi.org/10.1155/2024/4742401

\begin{abstract}
Opinion Dynamics is an interdisciplinary area of research. Disciplines of Psychology and Sociology have proposed models of how individuals form opinions and how social interactions influence this process. Socio-Physicists have interpreted the observed patterns in opinion formation in individuals as arising out of non-linearity in the underlying process and helped shape the models. Agent-based modeling has offered an excellent platform to study the Opinion Dynamics of large groups of interacting individuals. In this paper, we take recent models in opinion formation in individuals. We recast them to create a proper dynamical system and inject the idea of clock time into evolving individuals' opinions. Thus the time interval between two successive receipts of new information (i.e., the frequency of information receipts) by an individual becomes a factor that can be studied. In recent decades, social media has continuously shrunk time intervals between receipt of new information (i.e., increased the frequency of information receipts). The recast models are used to show that as the time interval between successive receipts of new information gets shorter and the number of individuals in one’s network becomes larger, the propensity for polarization of an individual increases. This explains how social media could have caused polarization. We use the word ‘polarization’ to mean an individual’s inability to hold a neutral opinion. A Polarization number based on sociological parameters is proposed. Critical values of the polarization number beyond which an individual is prone to polarization are identified. These critical values depend on psychological parameters. The reduced time intervals between receipt of new information and an increase in the size of groups that interact can push the Polarization number to approach and cross the critical value and could have played a crucial role in polarizing individuals and social groups. We also define the Extent of Polarization as the width of the region around neutral within which an individual is unable to have an opinion. Reported results are for values of model parameters found in the literature. Our findings offer an opportunity to adjust model parameters to align with empirical evidence. The models of opinion formation in individuals and the understanding arrived at in this study will help study Opinion Dynamics with all its nuances and details on large social networks using Agent-Based Modeling.
\end{abstract}
\textbf{Keywords:} Opinion Dynamics, Agent-Based Modeling, Polarization, Dynamical System, Polarization Number
\section*{Introduction}
Individuals have opinions on ‘things’ that concern them or ‘things’ they care about. ‘Things’ could refer to other individuals, places, ideas, issues, ideologies, or any of the other things. How does the opinion of an individual evolve? How do the opinions of a group of individuals evolve collectively? When and why do groups of individuals get polarized in their opinion? These and more complex, interesting, and important questions form research topics in Opinion Dynamics. Google Scholar finds 1,730 papers with the “Opinion Dynamics” phrase in their title and 19,800 anywhere in the article. Close to 30 percent of these have appeared in the last 2 years. This indicates the interest that this study is receiving. Researchers continue their efforts to improve models to better understand and explain polarization. For example, Gaitonde et al. \cite{gaitonde} discussed novel opinion update models, and Musco et al. \cite{musco2021quantify} explored new ways to quantify polarization. Several survey papers exist in Opinion Dynamics \cite{Noorazar_2020,opinion_dynamics_review2,lorenz2007continuous,urena2019review,anderson2019recent}. Despite this large volume of work, authors in \cite{leonard2021nonlinear} feel that "the existence of polarization is evident but explanations for polarization are piecemeal and not holistic."

A meta-study reported by Lorenz-Spreen, et al. \cite{lorenz-spreen2021} stands out having comprehensively reviewed 496 published articles in Opinion Dynamics. They report a link between digital media and polarization. Biondi et al. \cite{biondi2023dynamics} observed that widespread usage of online social networks and social media seems to amplify the already existing phenomenon of polarization. Lee \cite{lee2016} defined a polarization score and used that score to show how the polarization score of citizens of Hong Kong towards its Government increased by a factor of about 3 times over the period 2003 to 2014. The appearance of major social media platforms Facebook (2004), Twitter (2006), and WhatsApp (2009) correlates with this increasing polarization but without establishing any causation. Hunt Allcott, et al. \cite{allcott2020} reported interesting results of an experiment, where people were made to deactivate Facebook accounts four weeks before the 2018 US election that made them less polarized. This establishes some causation. Both these studies suffer from the samples used, e.g. restricted nationality, etc. yet in that limited sense they show that the appearance of digital media correlates and may have caused polarization. Bavel, et al. \cite{vanbavel2021} called for more work to establish if and how social media impacts political polarization. This provides the primary motivation for our work.
A large quantity of the work in Opinion Dynamics invokes some form of Agent-based modeling (ABM) to arrive at conclusions. Agent-based models (ABM) set rules for agents to interact with other agents in a social group (here the term agent stands for an individual). There are rules for updating an agent’s opinion based on the opinions of self and interacting agents. Distribution of opinions across agents at the start is defined. Iterations are set to roll allowing several rounds of interactions between agents, monitoring the group for signs of polarization. The iterations in ABM represent advancing of time. But in most ABMs there is no concept of how much time lapses from one iteration to the next. That time could be 1 minute, 1 hour, or 1 day but that has no impact on the outcome; the results will be the same. This is where our main contribution comes in. We attach a wall-clock time interval to iterations and can study what happens to an individual/agent who interacts with another agent once a day and compare that with what would have happened if the interaction had happened twice a day, etc. Social media’s contribution has been to shorten the time interval between two successive interactions (i.e. increase the frequency of information receipts). This makes it possible for us to study the impact of social media on opinion dynamics.
(Socio)Physics has shaped models that explain how individuals form their beliefs and opinions. The ‘Cusp Catastrophe’ surface \cite{strogatz1994nonlinear} is central to explaining the observed phenomena. The ‘Cusp Catastrophe’ surface is used in \cite{scheffer2022belief} to graphically, and qualitatively build arguments about ‘belief’ that non-linearly depends on evidence and ‘confirmation bias’. Maas, et al. \cite{van_der_mass} have modeled opinion formation using equations whose steady state solutions (a.k.a fixed points) form a ‘Cusp Catastrophe’ surface. They \cite{van_der_mass} trace the basis of their models to \cite{Abe_2017,stanley1987introduction}. They state 3 hypotheses and then test those hypotheses using Agent-based modeling \cite{van_der_mass}. We start from the models proposed by \cite{van_der_mass} and thereafter significantly differ from \cite{van_der_mass} in how the models are used. We first restate the models so that they form a proper dynamical system. We then introduce the idea of wall clock time. We further employ techniques from non-linear dynamics, like attractors and their stability to gain an understanding of an individual. We believe that a good understanding of individual behavior is needed before embedding such individuals in social networks to study group behavior.
Opinion Dynamics has become an interdisciplinary area of work. The present authors have exposure to Flight Dynamics, Agent-based modeling and Engineering Systems Design. We have always imagined humans as far more complex systems than any man-made system we have studied. So, ‘modeling any aspect of human behavior using one or two equations appears naive. But sometimes details do not seem to matter’ \cite{BALL20021}. We use the equations, as they are, knowing that they represent the most complex system and not fully understanding the complexity they have simplified. Yet we go ahead to draw precise conclusions. We believe our findings can be used to fine-tune parameters in the opinion formation model using Parameter Identification techniques or in designing experiments to measure the parameters. Fitting catastrophe models to data has already been reported in this domain \cite{grasman2010fitting,van2003sudden} and is routinely carried out in the larger context of dynamics and introduced in several texts \cite{isermann2011identification}
We write this paper using standard terms in Opinion Dynamics but may use terms in non-linear dynamics where necessary. We hope that it shall communicate well with all those who are from other expert areas.
In Section 2 we describe the basic terminology in Opinion Dynamics, the variables in the problem and briefly introduce ideas in non-linear dynamics. In Section 3 we discuss the models and restate them as proper dynamical systems. In Section 4 we use the models to gain insights into individuals' behavior. This focuses on studying the impact of the frequency of arrival of new information on an individual's propensity to get polarized. This is where we propose a Polarization number and define its critical values. In Section 5 we conclude.
\section*{2. Opinion Dynamics – Terms and what they mean }
We shall first introduce the models proposed by \cite{van_der_mass}. We use the same notations where possible but deviate in defining the parameters so they are more meaningful while interpreting the resulting behavior.
\subsection*{2.1 Attitude}
An Attitude object can be an issue, an ideology, an idea, an individual, etc. Attitude in respect of an attitude object is formed over time and is influenced by learning. Learning occurs when an individual comes into contact with new information about the attitude object.
An individual’s attitude towards an attitude object has 3 components, viz. Affect, Behavior and Cognition \cite{daffin2021principles}. Opinion is one aspect of the cognition component of attitude and the present study focuses only on opinion.
\subsection*{2.2 Opinion}
Opinion, $O$, is a judgment formed about an attitude object. In this paper Opinion about an attitude object is taken as continuous and bounded $ -1 \leq O \leq +1 $. Here $-1$ indicates a stand extreme to one side and $+1$ extreme stand to the other side, and $0$ indicates a neutral stand. Other contexts can use a simpler representation of $O$, e.g. binary $[0, 1]$. Yet other contexts can demand more complex representations of $O$.
\subsection*{2.3 Attention}
“Attention is mind taking possession of one out of several possible attitude objects. It implies withdrawal from some things to deal effectively with others” \cite{james1890principles}. We represent attention by $A$, where $A \geq 0$. Upper bound of A is left to the models to reveal.
\subsection*{2.4 Information}
Information is a piece of fact provided or learned about an attitude object. We represent a piece of information by $i$. Note that, $ -1 \leq i \leq +1 $. When $ i = -1 $, it fully validates one extreme stand; when i=+1, it fully validates the other extreme stand.
What is presented to an individual as fact can be fact or fake. This important distinction is not made in the present study as it does not influence its findings. Such distinction and many other nuances can be used to understand finer details of polarization like the distribution of converged opinions of individuals, how many individuals are in each polarized group, how far apart the opinions of the two groups are, etc., using Agents-based modeling.
At any instant in time, an individual has a particular level of information about an attitude object which is the cumulative result of all pieces of information received/learned till then. We represent this by $I$. When a new piece of information, $i$, reaches an individual, they update their information, $I$.
Note that information is provided to an individual by other individuals or by media (print, TV or social media). Information can also be purposefully learned by an individual.
\subsection*{2.5 Dynamics}
Dynamics began as a branch of mechanics concerned with the motion of bodies under the action of forces. A broader view of dynamics includes anything that changes with time. Since the opinion of individuals changes with time, its study is referred to as Opinion Dynamics.
In dynamics, one has to identify the states. Once the states are known the equations that govern their evolution with time must be arrived at by modeling. The evolution of states can depend on the value of states and the values of prime movers (i.e. what causes their change). It can also depend on a set of parameters.
The following are of interest while studying a dynamical system. More detailed descriptions can be found in textbooks on dynamics, e.g. \cite{strogatz1994nonlinear}.
\begin{enumerate}
\def\labelenumi{\arabic{enumi}.}
\item Simulation. For the desired set of initial values of the states, simulation can evolve the states for as long a time as needed. Depending on the nature of the model and equations, an appropriate solution strategy has to be used.
\item Attractors. For any arbitrary initial conditions if the simulation is carried out for a long enough time we shall approach an attractor. An attractor can be a steady state (a.k.a fixed point or equilibrium), a periodic attractor or a chaotic attractor. This study shall be interested only in the first two attractors. A steady state is where the rate of change in the states is zero. Periodic attractors endlessly retrace the same trajectory in state space with a periodicity. There are techniques to directly identify attractors of a dynamical system than through simulation.
\item Stability. This property is what distinguishes an attractor from a repeller. If a dynamical system is slightly perturbed from an attractor it shall return to the attractor. Dynamical systems can have fixed points or periodic solutions that are repellers or unstable. The system can be carefully placed at these unstable solutions but shall get pushed away from it by the smallest disturbances. There are techniques to investigate a fixed point or a periodic solution for its stability.
A dynamical system, when left to itself shall be found at one of its attractors but not at a repeller. This interesting property shall be invoked later.
\item Basin of Attraction: An attractor shall attract all trajectories in some neighborhood to it; hence the name attractor. All attractors have a basin of attraction. If the system is at states within the basin of attraction of an attractor it will get attracted to it.
\end{enumerate}
\section*{3. Opinion Dynamics: The Models}
There are two states in the models proposed by \cite{van_der_mass}, opinion $(O)$ and attention $(A)$. The forcing function (prime mover) is Information $(I)$. In addition, there are parameters too.
There are two equations governing the evolution of the two states. The first of these is presented by \cite{van_der_mass} as a differential equation.
\begin{equation} \label{eq:1}
\frac{dO}{dt} = \  - \left( O^{3} - \left\{ A - A_{crit} \right\}\ O - I \right)
\end{equation}
We have not retained the Wiener noise term from \cite{van_der_mass} as this study aims to understand the dynamics of opinion formation of an individual under ideal conditions. $ A_{crit} $ is a parameter and its role shall become clear later. It is the negative of the parameter $ A^{min} $ in \cite{van_der_mass}.
The solution that emerges from above has to be normalized to be bounded within [-1, +1]. The scaling factor and bias, if any are required, shall be decided once the solution emerges.
The second equation governing attention, A is given by \cite{van_der_mass} as.
\begin{equation} \label{eq:2}
\frac{dA}{dt} =-\frac{2k}{N^{2}} A + k \left(A_{max} - A \right)
\end{equation}
The first term represents the decay of attention. The second term represents a spike in attention each time new information is received ($u=1$ at the instant new information is received and $u=0$ at all other times).
The term $dA$ on the left-hand side gives it the appearance of a differential equation, but there is no $dt$. The rate of change in $A$ is supposed to be represented by $d_{A}$. So we relate $d_{A}$ to $dt$ by introducing the parameter $k$, $d_{A} = k dt$. As per results reported in \cite{van_der_mass} the second term is strictly not a spike in $\frac{dA}{dt}$ but rather in $A$. i.e. it is not $\Delta\left( \frac{dA}{dt} \right)$ but $\mathrm{\Delta}A$. So we represent it accordingly. These changes recast it as a proper dynamical system. We further replace the parameter $ A^{*} $ by $ A_{max} $ = 2 $ A^{*} $. The subscript 'max' is to suggest that this will be the maximum value of A.
\begin{equation}
\frac{dA}{dt} =-\frac{2k}{N^{2}} A + k \left(A_{max} - A \right)
\end{equation}
where $ \Delta A $ is a spike in $A$ when new information is received and $b=\frac{2k}{N^{2}}$. The time represented by 't' in equations [1] and [2] is in seconds. The parameter $k$ is akin to inertia in dynamics and to be more precise inverse of inertia. In response to a forcing function, if $k$ is small, $A$ shall change gradually and shall change by smaller magnitude and vice versa.
\section*{4. Opinion Dynamics: Individual Behavior Based on Models}
Before we try to understand the dynamics of a group of individuals, it is good to understand individual behavior. An individual has two states, opinion $O$ and attention $A$. There is only one driving force, Information $I$. Equations \ref{eq:1} and \ref{eq:2} govern their dynamics:
Steady-state (a.k.a fixed point or equilibrium) can be found by equating 
$\frac{dO}{dt}\ \&\ \frac{dA}{dt}$ to zero and solving for values of
$O$ \& $A$. For each value of $I$ there will be associated steady states.
Note: Each $I$ can have more than one steady state associated with it.
Each steady state can then be investigated for its stability to establish if it is an attractor or repeller.
\subsection*{4.1 Dynamics of Attention}
The equation \ref{eq:2} governing $A$ is independent of $O$; i.e. it is decoupled. Hence
its evolution can be studied in isolation. The steady state of $A$, in the
absence of any new information reaching, can be found by
setting $\frac{dA}{dt} = 0$. Inspection of equation \ref{eq:2} will confirm that $A =0$
is the steady state. Equation \ref{eq:2} can be integrated analytically to get:
\begin{equation} \label{eq:3}
A_{t} = A_{0}\ e^{- b\ t}
\end{equation}
The analytical solution equation \ref{eq:3} shows that no matter the value of $A$
at time $t=0$, it shall eventually tend to the steady state $A_{t} = 0$ as $t$
tends to $\infty$. So, the steady state, $A=0$, is an attractor and its basin of
attraction is all of the space; i.e. it will attract all states to it.
Dynamics of $A$ has another part, which is the spike in $A$ when new information is received. The arrival of new information causes the spike, but the value of the spike does not depend on the information. We exploit this important fact and model only the time instance when interaction with another agent happens (i.e. time when new information is received) and do not find the need to model the information exchange and the update logic. This makes it possible to do away with Agent-based Modeling in our approach.
\begin{equation}
\Delta A = k \left(A_{max} - A \right)
\end{equation}
$k=0.2$ and $A_{max} = 2$ are the values assigned to the parameters by
\cite{van_der_mass}. It can be seen that for $A \leq A_{max}$, $ A + \Delta A \leq A_{max}$, i.e. starting from $A \leq A_{max}$, $A$ shall not grow beyond $A_{max}$ 
We have seen that the attention decays continuously and there is a spike each time information is received. This raises an interesting question. At a given periodicity, $\tau$ of information receipt how does $A$ vary? The inverse of $\tau$ is the frequency of information receipts. Henceforth we shall use $\tau$ as the variable. One may wonder why periodicity in information arrival is of interest. Typically there is a rough periodicity in receipt of information in an individual's life. Some individuals may be meeting friends every day in the evening, for some it may be only on weekends, some may be reading print media (newspaper) delivered daily or maybe watching some specific TV news bulletin at specific hours each day, or checking WhatsApp messages every 2 hours, etc. Though these may not happen with precise periodicity, this idealized condition can give important insights, as we shall soon see.
We explore the behavior of $A$ for periodicity of information arrival, $\tau$. Let $ A_{0} $ be the value of $A$ at time $t=0$. Attention, $A$ decays to $A_{0}e^{- b\tau}$ and then spikes by $\Delta A$ to take a value $A_{\tau}$ at time $\tau$;
\begin{figure}
\centering
\includegraphics[width=1\linewidth]{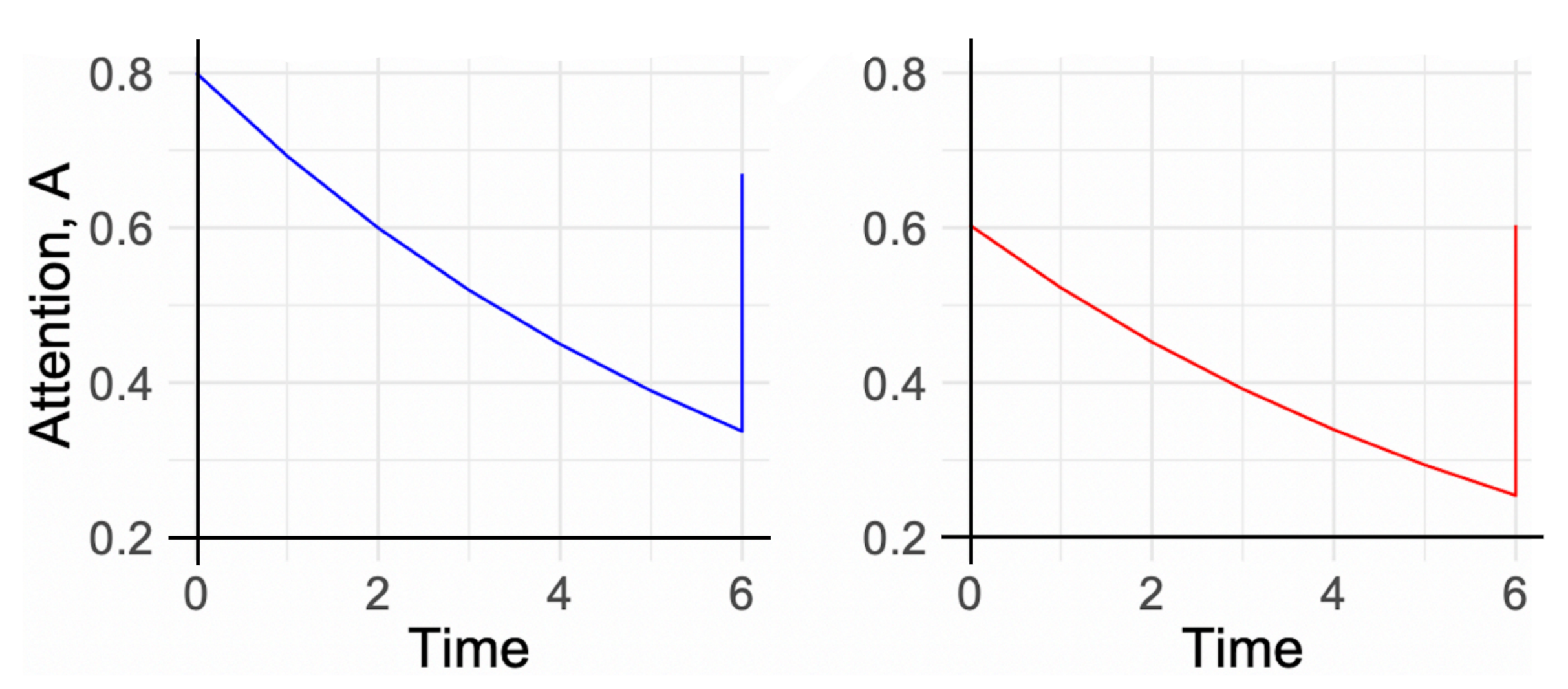}
\caption{At time $t=0$, there exists a specific initial value of $A$ that will decay with time and when new information arrives after time $\tau$, spike back to the same initial value. For $\tau=6$ hours, we illustrate the time history both for an arbitrary initial value and the specific initial value.}
\label{fig:tau1}
\end{figure}
$A_{\tau} = A_{0}\ e^{- b\tau} + \Delta A$
$= A_{0}{\ e}^{- b\tau} + k\left( A_{max} - \ {A_{0}\ e}^{- b\tau} \right)$
$= \ {A_{0}\ (1 - k)\ e}^{- b\tau} + kA_{max}$
For any arbitrary value of $A_{0}$ we shall not have
$A_{\tau} = A_{0}$ \{Figure \ref{fig:tau1}\}. So, we set $A_{0} = A_{\tau}$ ; and solve for $A_{0}$ and call it $A_{0\tau}$
$A_{0}=A_{\tau}{= A_{0}\ (1 - k)\ e}^{- b\tau} + kA_{max}$  
\begin{equation} \label{eq:5}
A_{0\tau} = \frac{kA_{max}}{1 - {\ (1 - k)\ e}^{- b\tau}} 
\end{equation}
If at $t=0$, $A = A_{0\tau}$ and new information is received at
periodicity of $\tau$, then we shall have a solution that is periodic,
with periodicity, $\tau$ \{Figure \ref{fig:tau1}\}. The periodic solution for any $\tau$ is bounded by
$A_{L\tau}$ and $A_{U\tau}$; whose values for any $\tau$ can be computed using:
\begin{equation}
A_{U\tau}=A_{0\tau} =\frac{kA_{max}}{1 - {\ (1 - k)\ e}^{- b\tau}} 
\end{equation}
\begin{equation}
A_{L\tau} = \ A_{0\tau}e^{- b\tau}\  = \frac{kA_{max}\ e^{- b\tau}}{1 - {\ (1 - k)\ e}^{- b\tau}}\ 
\end{equation}
For $\tau=6$ hours $A_{U6}= 0.6034805$ and $A_{L6}=0.2543506$. 
\newline
We now define the Polarization number, P as 
\begin{equation} \label{eq:pol}
P = \frac{\tau}{N^{2}}
\end{equation}
Present authors are familiar with numbers like Reynolds number, and Mach
number in fluid flows. When these numbers cross a critical value the
nature of fluid flow changes. We shall similarly show that when
Polarization number crosses a critical value the nature of opinion
formation in individuals changes. We rewrite equations \ref{eq:5}, \& \ref{eq:6} in terms of $P$,
\begin{equation}
A_{UP} = \frac{kA_{max}}{1 - {\ (1 - k)\ e}^{- 2kP}}\ 
\end{equation}
\begin{equation}
A_{LP} = \frac{kA_{max}\ e^{- 2kP}}{1 - {\ (1 - k)\ e}^{- 2kP}}\ 
\end{equation}
Figure \ref{fig:image3} shows the variation of $A_{UP}$ and $A_{LP}$ with
$P$. Low $P$ (i.e low $\tau$ or high $N$) can keep the attention high. This is nothing new. Zajonc \cite{zajonc1968attitudinal} had shown, as early as 1968 that mere repeated exposure of an individual to a stimulus object enhances his attitude toward it. But here the relationship is quantified.
It may be noted that $A$ takes its
upper bound value of $A_{max}$ (which is 2 in this case) as $P \rightarrow 0$. This is consistent with what we stated in Section 4.1.
\begin{figure}
\centering
\includegraphics[width=1\linewidth]{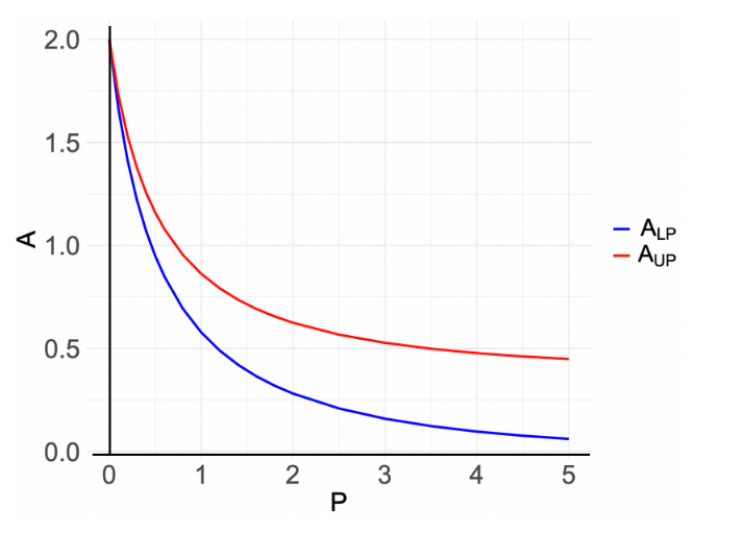}
\caption{For a given polarization number (P), Attention ($A$) oscillates between $A_{UP}$ and $A_{LP}$. Both values increase with decreasing polarization number ($P$) and eventually converge to $A_{max}$, in this case, 2.}
\label{fig:image3}
\end{figure}
Now let us turn our attention to another interesting behavior. For $\tau= 6$ hours consider two cases, $N=100$ and $500$. Set $A$ to some arbitrary value at $t=0$, not equal to $A_{0\tau}$. Say, $A_{0}=1$. Figure \ref{fig:image4} shows that in both cases the solutions quickly settle down to respective periodic solutions. This suggests that these two periodic solutions act like attractors. So we investigate periodic solutions to check if they are attractors.
\begin{figure}
\centering
\includegraphics[width=1\linewidth]{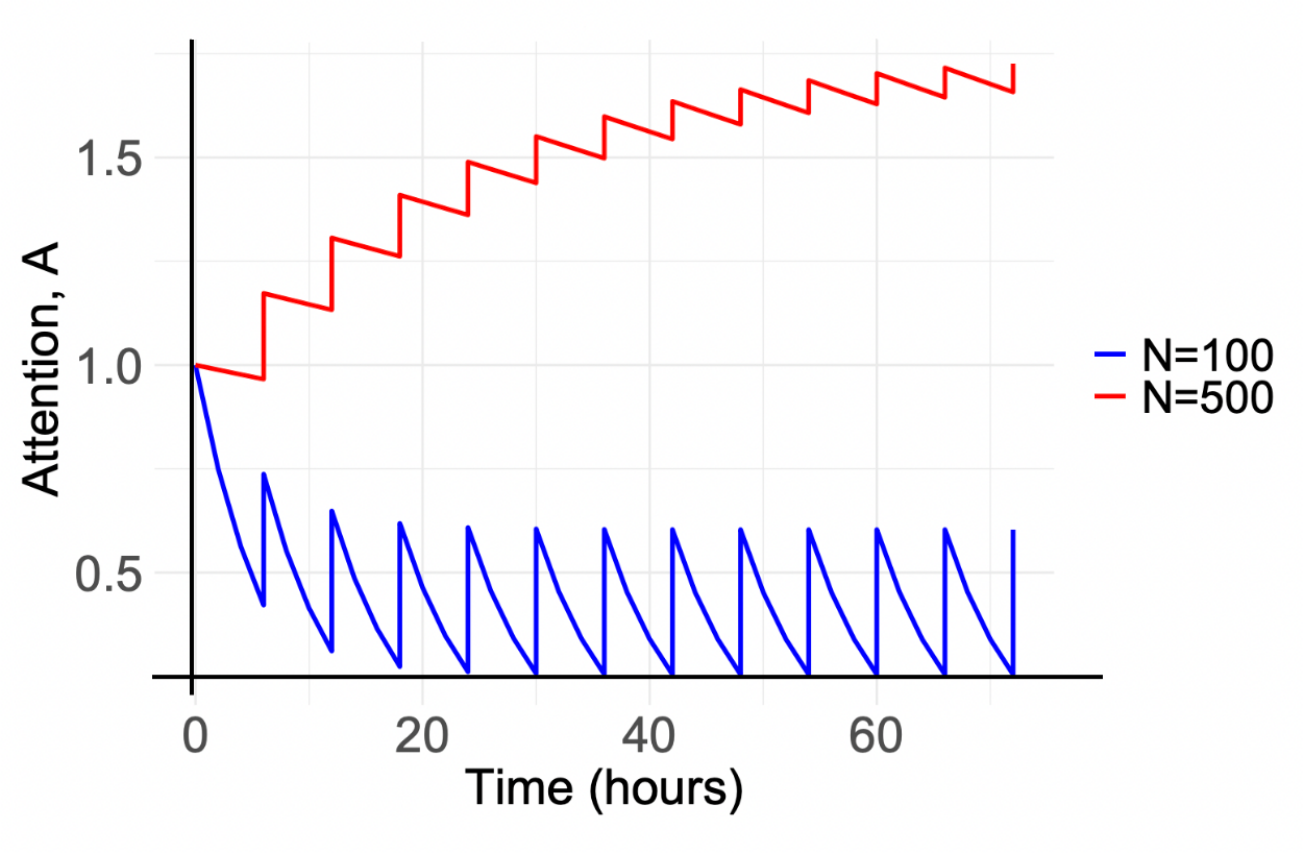}
\caption{For a given periodicity of information arrival ($\tau$), starting from any arbitrary value, attention ($A$) quickly converges to the corresponding periodic solution. For $\tau=6$ hours, two cases of $N$ are illustrated.}
\label{fig:image4}
\end{figure}
Consider a time period $\tau$ and associated $A_{0\tau}$. If the starting value of $A_{0}=A_{0\tau}$ ; we shall have as per definition;
\begin{equation*} 
A_{\tau} = A_{0\tau} e^{-b\tau} + k \{ A_{max} - A_{0\tau} e^{-b\tau} \} = A_{0\tau}
\end{equation*}
If the starting value of $A$ is not $A_{0\tau}$ but $A_{0\tau}+a$, where $a$ is such that, 
$0 \leq A_{0\tau} + a \leq A_{max}$;
\[
A_{\tau} = (A_{0\tau}+a)e^{-b\tau} + k \{ A_{max} - (A_{0\tau}+a)e^{-b\tau} \}
\]
\[
= A_{0\tau} e^{-b\tau} + k \{A_{max} - A_{0\tau} e^{-b\tau} \} + a(1- k) e^{-b\tau}
\]
\[
= A_{0\tau} + a (1- k) e^{-b\tau}
\]

For $0 < k < 1$, $0 < (1-k) < 1$ 
we shall have ; 
$a (1- k) e^{-b\tau} < a$.
We started at $A_{0\tau}+a$ and after time $\tau$ ended up at $A_{0\tau} + a (1- k) e^{-b\tau}$; i.e. closer to $A_{\tau}$ than when started. After $n$ cycles (i.e. time $n\tau$)
the trajectory shall end up at $A(n\tau) = A_{0\tau} + a (1- k)^{n} e^{-nb\tau}$. With each cycle it is coming closer to $A_{\tau}$ and shall finally tend to
$A_{0\tau}$; i.e get attracted to the periodic solution corresponding to $\tau$
\subsection*{4.2 Dynamics of Opinion}
\begin{figure}
\centering
\includegraphics[width=1\linewidth]{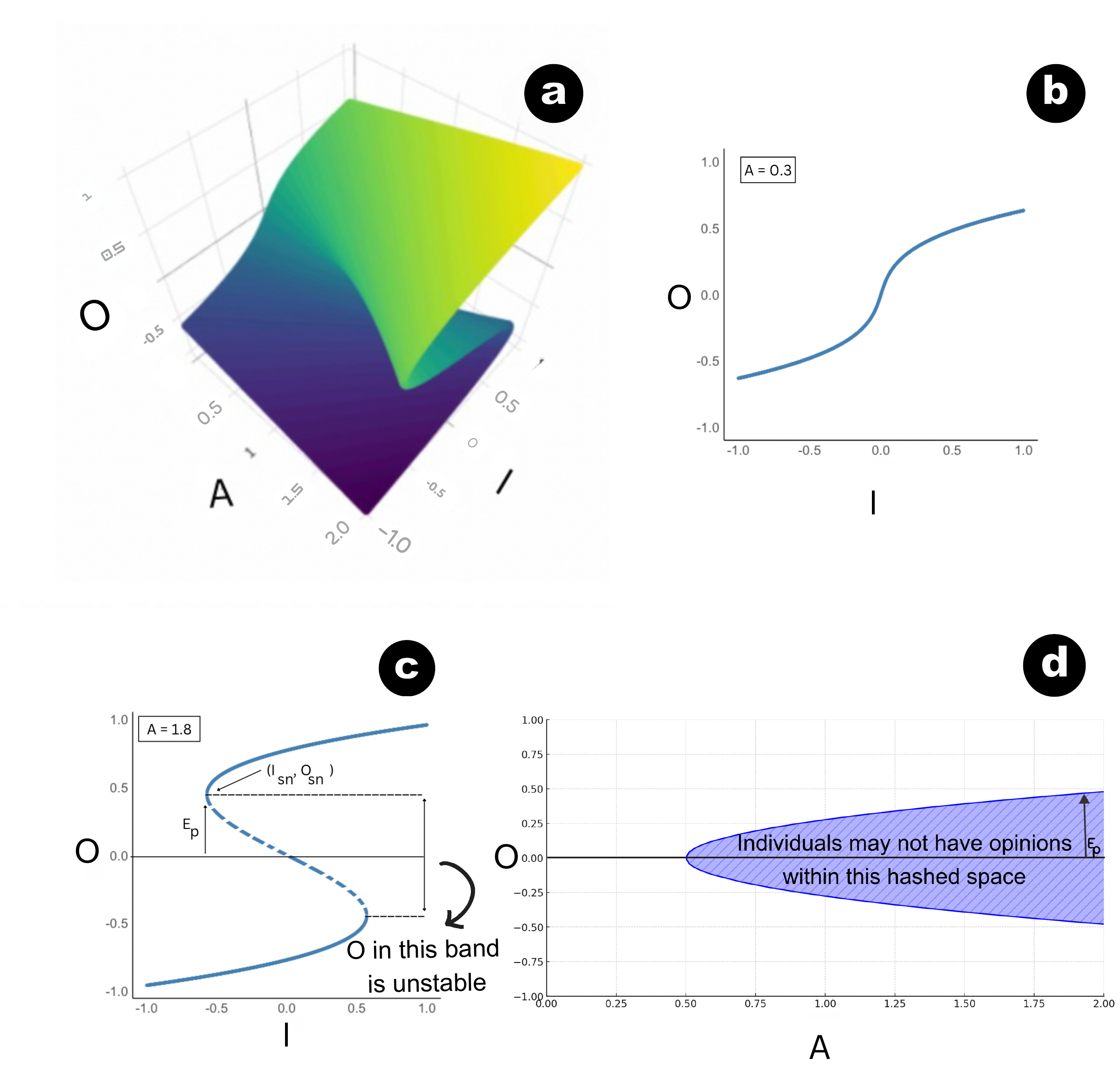}
\caption{Parameter values: $A_{\text{crit}}=0.5$, $A_{\text{max}}=2.0$, $k=0.2$. The Cusp Catastrophic surface for $O$ is scaled such that  $-1 \leq O \leq 1$ for $0 \leq A \leq A_{\text{max}}$  and  $-1 \leq I \leq 1$. No bias was required and the scaling factor used is inverse of max($A$) = 1.475687. 
(A) The Cusp Catastrophe Surface. The fold begins at $A=A_{\text{crit}}=0.5$, and the surface extends until $A=A_{\text{max}}=2.0$. (B) The cross-section at $A=0.3$ reveals that $O$  is stable over its full range of values across neutral. (C) The cross-section at $A=1.8$ reveals no stable opinions in the range $-O_{\text{sn}} \leq O \leq O_{\text{sn}}$. Individuals can either have opinions in the band $-1 \leq O \leq -O_{\text{sn}}$ or $O_{\text{sn}} \leq O \leq 1$. The extent of polarization at $A=1.8$ is $E_p=O_{\text{sn}}=0.46$. (D) The shaded region indicates the area where no stable opinions are possible. The extent of polarization, $E_p$, is the semi-height of the shaded region. $E_p = 0$ for $A \leq A_{\text{crit}}$ and increases thereafter with $A$.}
\label{fig:image5}
\end{figure}

Let us now turn our attention to equation \ref{eq:1} which governs the dynamics of
Opinion. We shall treat $I$ as a constant (no receipt of new
information).
\begin{equation*}
\frac{dO}{dt} = \  - \left( O^{3} - \left\{ A - A_{crit} \right\}\ O - I \right)
\end{equation*}
Steady states (a.k.a. fixed points) can be obtained by setting
$\frac{dO}{dt} = 0$;
\begin{equation} \label{eq:13}
O^{3} - \left\{ A - A_{crit} \right\}\ O - I = 0
\end{equation}
Let us take $A_{crit} = 0.5$ as in \cite{van_der_mass}. We shall see later
what $A_{crit}$ represents.
This is a non-linear cubic equation in $O$ and shall have three roots. Real roots 
represent steady states. For any pair of values \{$A, I$\} equation \ref{eq:13}
can either have one real root and a complex pair or three real roots.
Let us consider our region of interest, $0 \leq A \leq 2$ and $-1 \leq I \leq +1$.
Figure \ref{fig:image5}a is a plot of the surface $O(A,I)$ that represents the steady
states. This is the typical ‘Cusp Catastrophe’ surface \cite{strogatz1994nonlinear} .
The region where we have 3 real roots shows up as a fold in the surface.
The fold is for $A > A_{crit}$. The parameter
$A_{crit}$ controls the start of the fold.
Let us represent any point on this surface as ($O^{SS}$,
$A^{SS}$, $I^{SS}$). This is a steady state.
Let us investigate the outcome of a small disturbance from that steady
state to ($O^{SS}+o$, $A^{SS}$,
$I^{SS}$). If we substitute this in the differential
equation \ref{eq:1} and note that
$({O^{SS}}^{3} - \left\{ A^{SS} - A_{crit} \right\}\ O^{SS} - I^{SS} = 0,$
and set higher order terms in $o$, i.e. $o^{3}$ and $o^{2}$ to zero, we
get
\begin{equation*}
\frac{do}{dt}\frac{1}{o} = - \left( 3{O^{SS}}^{2} - (A^{SS} - A_{crit}) \right)
\end{equation*}

The condition for stability (or to be an attractor) is
$\frac{do}{dt}\frac{1}{o} < 0$ ; which implies that if $o > 0$, $\frac{do}{dt} < 0$ and if $o < 0$,
$\frac{do}{dt} > 0$. So, if we disturb the system from a steady state it
tends to cancel that disturbance.
Condition for stability;
\begin{equation} \label{eq:9}
- \left( 3{O^{SS}}^{2} - (A^{SS} - A_{crit}) \right) < 0
\end{equation}

If $O$ is at an unstable steady state, any small disturbance that can result from any source will result in $O$ moving away from the steady state and settling at a stable steady state. Hence we may not find any individual holding an $O$ corresponding to any of the unstable steady states. For a more detailed discussion on stable/unstable solutions, bifurcations, jump phenomena, hysteresis etc. refer to the book on non-linear dynamics \cite{strogatz1994nonlinear}.
Figure \ref{fig:image5}$b$ is a cross-section of the Cusp Catastrophe surface at $A < A_{crit}$. Figure \ref{fig:image5}$c$ is a cross-section at $A > A_{crit}$. Solid lines represent stable
steady states (attractors) and dashed lines represent unstable steady states (repellers). For values of 
$A \leq A_{crit}$, there is no fold and $O$ changes in a continuous manner over the entire range $-1 \leq I \leq +1$ and is stable all along. An individual with 
$A \leq A_{crit}$, can have any value of $O$ across the neutral region Figure \ref{fig:image5}b. For values of $A > A_{crit}$, there is a fold and there is a band of values of $O$ near neutral that have only unstable steady states. We may not find any individual holding $O$ within this band. $O$ suddenly jumps (discontinuously) across the band of unstable steady states at two critical values $\pm I_{sn}$. Corresponding values of Opinion are $\mp O_{sn}$ Figure \ref{fig:image5}c. The subscript $_{sn}$ is to suggest that they are Saddle Node bifurcation points.
We use this to define the Extent of Polarization of an individual as, $E_{P} = |O_{sn}|$. Thus an individual cannot have an opinion in the band
$-E_{P} < O < E_{P}$. This is the band that separates two opposing polarized individuals. 
Note that an individual can have an opinion only in one of the two bands, $-1 \leq O \leq -E_{P}$ or $E_{P} \leq O \leq 1$.
For $A \leq A_{crit}$ we have $E_{P}=0$. As 
$A$ crosses $A_{crit}$, and increases to its limiting value 2, $E_{P}$ increases monotonically from zero. Variation of $E_{P}$ with $A$ is in \{Figure \ref{fig:image5}d\}
\subsection*{4.3 Some interesting observations}
We have seen that for any given Polarization number, $P$ the attention, $A$ settles down to a periodic solution corresponding to that Polarization number, $P$. This periodic solution has $A$ within the bounds $A_{LP}$ and $A_{UP}$. Consider two limiting values of $P$ denoted by $P_{1}$ and $P_{2}$. If for $P \geq P_{1}$, $A_{UP}  \leq A_{crit}$; then $A$ shall always be in the region where there is no fold, and $E_{P}=0$. i.e the extent of polarization is zero. Similarly if for $P \leq P_{2}$, $A_{LP} \geq A_{crit}$ then $A$ shall always be in the region with fold, i.e. $E_{P} \geq 0$ i.e. prone to polarization. $P_{1}$ and $P_{2}$ can be estimated using equation \ref{eq:7_1} and  \ref{eq:7} as follows:
Solving for $P$ from
\begin{equation}
A_{UP} = \frac{kA_{max}}{1 - {\ (1 - k)\ e}^{- 2kP}} = A_{crit}
\end{equation}
gives $P_{1}$
\begin{equation} \label{eq10}
P_{1} = - \frac{1}{2k}\ln\left\{ \frac{1 - \left( \frac{kA_{max}}{A_{crit}} \right)}{1 - k} \right\}
\end{equation}
Solving for $P$ from
\begin{equation}
A_{LP} = \frac{kA_{max}e^{- 2kP}}{1 - {\ (1 - k)\ e}^{- 2kP}} = A_{crit}
\end{equation}
gives $P_{2}$
\begin{equation} \label{eq11}
P_{2} = - \frac{1}{2k}\ln\left\{ \frac{1}{\frac{kA_{max}}{A_{crit}} + 1 - k} \right\}
\end{equation}
For $k=0.2$, $A_{crit} = 0.5$ and $A_{max}=2$ we have
$P_{1} = 3.466$  and $P_{2} = 1.175$,  {Figure 5}.
\begin{figure}
\centering
\includegraphics[width=1\linewidth]{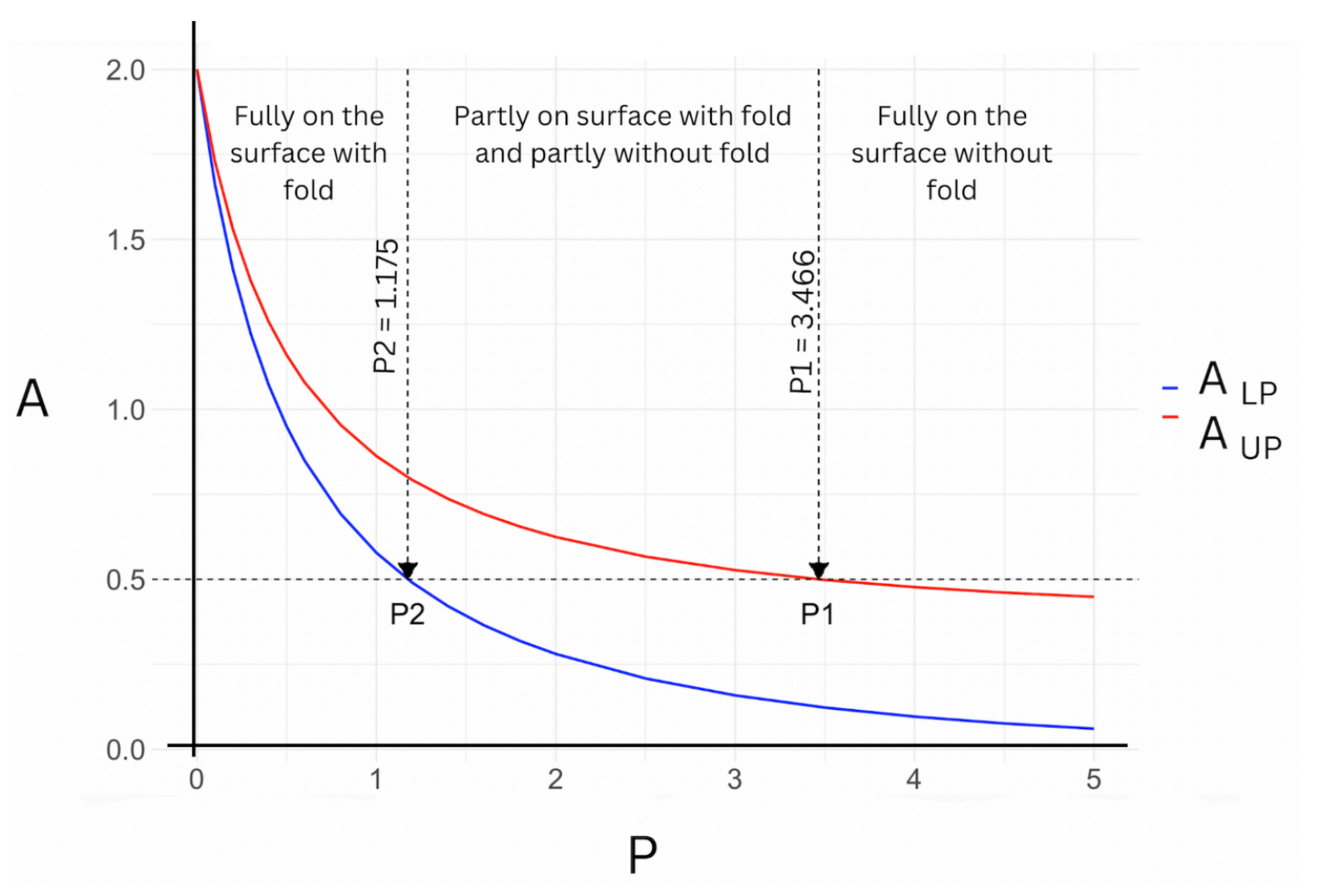}
\caption{Critical values of polarization numbers, $P_1$ and $P_2$, bracket the transition region. On one side lies the region where the extent of polarization, $E_p$= $0$ and the other side where $E_p > 0$ respectively.}
\label{fig:image6}
\end{figure}
Individuals with a Polarization number greater than 3.466 have $E_{P}=0$ and can have Opinions over its range across neutral. Individuals with a Polarization number
less than 1.175 have $E_{P} > 0$ and can have
opinions only in one of the two polarized regions,
$O < -E_{P}$ or $O > E_{P}$. $P_{1}$ marks the threshold of
polarization and $P_{2}$ marks its completion. Individuals
with a Polarization number in the range of $1.175 < P < 3.466$ can exhibit a mixed behavior. This is an important finding.
Let us take a fresh look at the model. Time, $t$ is the independent variable. $k$, $A_{max}$ \& $A_{crit}$ are parameters in the model that are psychological. $\tau$ \& $N$ are more sociological variables. The Polarization number $P = \frac{\tau}{N^{2}}$ combines the two sociological variables into one composite variable. Note that $P$ is not a non-dimensional parameter. Critical values of Polarization number beyond which an individual has a propensity to polarization, $P_{1}$ \& $P_{2}$, are functions of only psychological parameters, $k$, $A_{max}$, $A_{crit}$. We can map these boundaries in terms of $N$ and $\tau$ (Figure \ref{fig:image7}).
\begin{figure}
\centering
\includegraphics[width=0.5\textwidth]{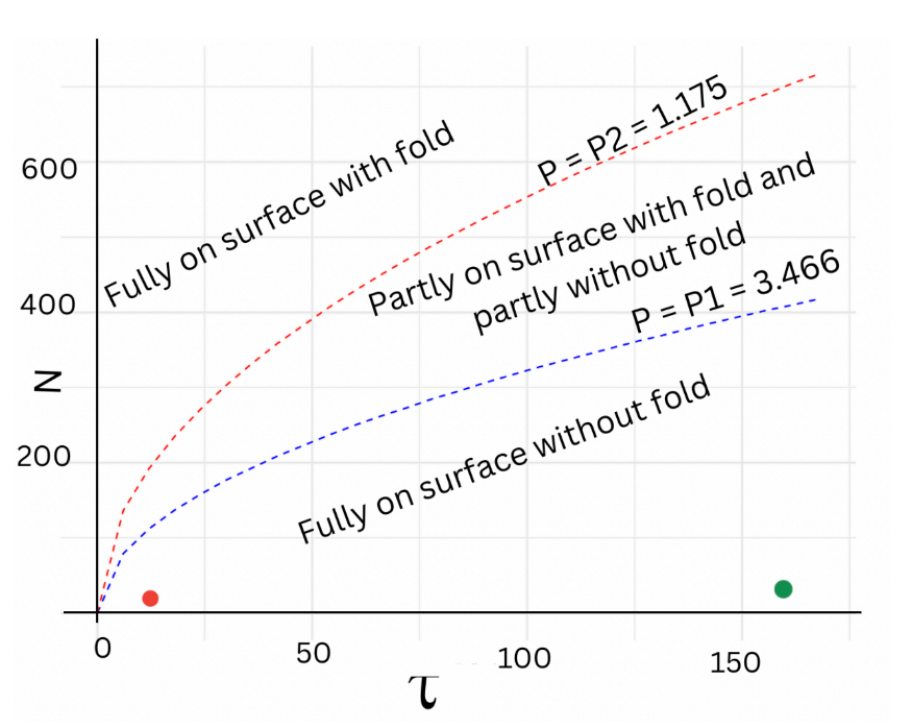}
\caption{The transition region, as well as non-polarized and polarized regions, are depicted on the $\tau$ vs. $N$ diagram.}
\label{fig:image7}
\end{figure}
One of the authors currently at the age of 75 recalls past years when contact with others was entirely by landline telephones with $\tau \approx 6-7$ days $\approx 168$ hrs and with group sizes of 5 to 10 and presently with $\tau \approx 0.5$ days $\approx 12$ hrs and with group sizes of 15 to 20. These are marked in green and red dots (Figure \ref{fig:image7}).
We may recall that the increase in political polarization from 2003 to 2014 in Hong Kong correlates with the arrival of social media \cite{lee2016} and that abstaining from social media by some Americans reduced their polarization \cite{allcott2020}. These studies tend to explain polarization based on the content on social media. Our study shows that the underlying reason is not the content itself but the frequency at which new content becomes available and is accessed. The content only decides the side to which the individual gets polarized. We agree with \cite{vanbavel2021} that research so far has not established how social media impacts polarization and we agree with \cite{lorenz-spreen2021} that explanations offered for polarization so far are piecemeal and not holistic. Our explanation of polarization differs significantly from what exists. Our explanation of polarization is argued with rigor starting with the models that are borrowed.
\begin{figure}
\centering
\includegraphics[width=1\textwidth]{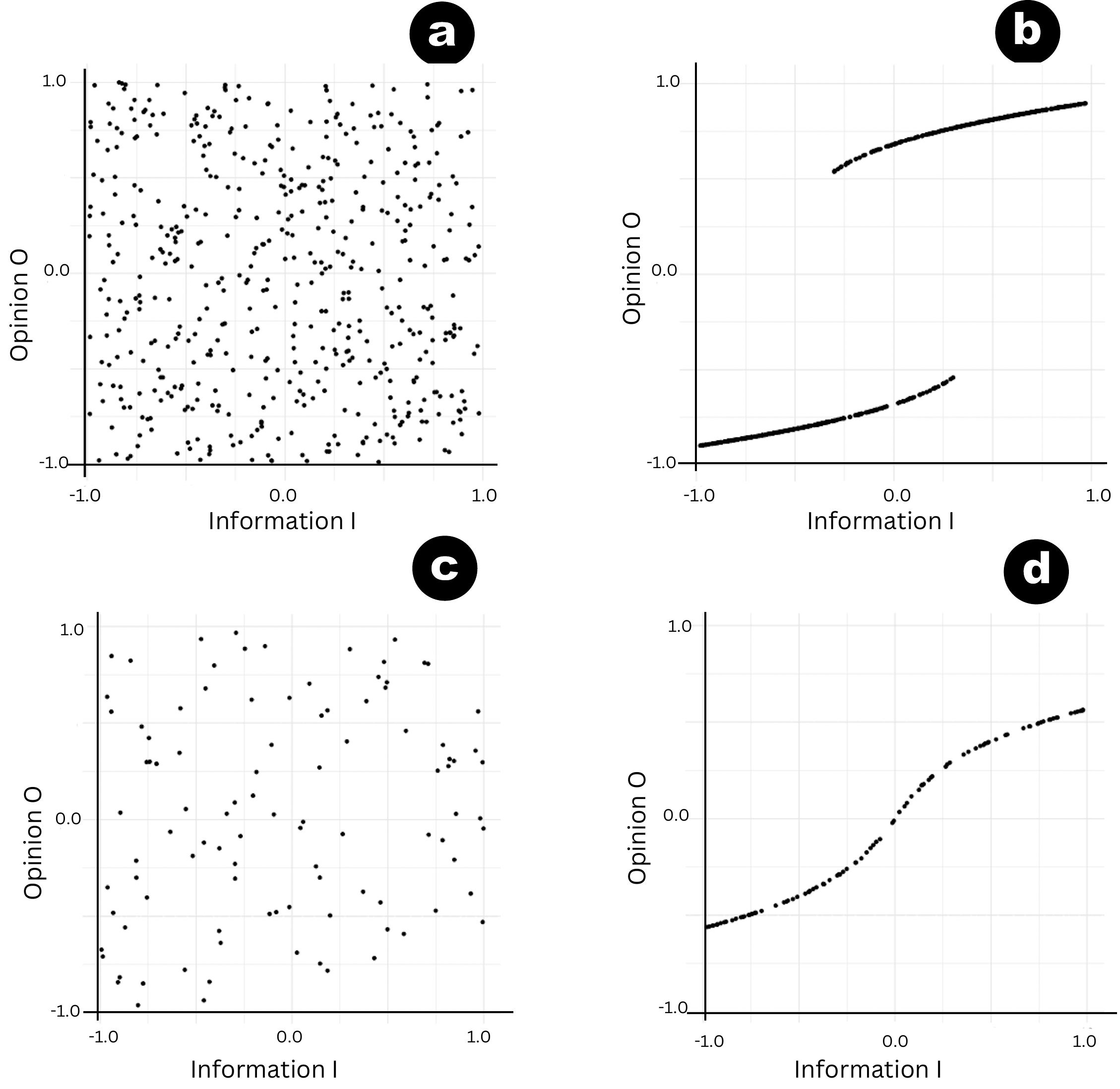}
\caption{Results from Agent Based Modeling. Figures (a) and (b) depict the dynamics of 500 agents ($N=500$) who interact every 5 hours ($\tau=5$). (a) $I$ and $O$ of agents are uniformly randomly initialized. (b) After a series of interactions, no agent retains an opinion within a neutral band. This configuration, $N=500, \tau=5$, corresponds to $P=0.072 < P_{2}$, indicating polarization. Figures (c) and (d) depict the dynamics of 100 agents ($N=100$) interacting every 100 hours ($\tau=100$). (c) $I$ and $O$ of agents are uniformly randomly initialized. (d) After a series of interactions, agent opinions are distributed across the full range ($-1 < O < +1$). For $N=100, \tau=100$, the resulting $P=36 > P_{1}$ suggests no polarization. Animated versions of these simulations, showing values of I, O for each agent at periodic intervals, are available on \href{https://github.com/shaja/PolarisationNumber}{this GitHub page}.}
\label{fig:image_last}
\end{figure}
We have studied the ideal case where new information arrival is periodic and happens at precise time intervals of $\tau$. In real life, such clockwork precision is unlikely. So we use simulation to study more realistic scenarios. Consider the ideal case corresponding to $\tau = 6$ hours and $N=100$. We have seen that starting from any arbitrary value of $A$, after a few cycles the solution is periodic and bounded by $A_{U,\tau=6}= 0.6034805$ and $A_{L,\tau=6}=0.2543506$. We shall refer to these as base values. Now, consider a case where the time lapsed between two successive arrivals of new information is $D\tau = \tau + e(0, w\tau)$. Here $e$ is a normally distributed random number with mean $\mu =0$ and standard deviation $\sigma = w\tau$. Simulations for $\tau = 6$ hours and $N=100$ and realistic $w$ values show $A$ has its upper and lower values in each cycle close to their respective base upper and lower bound values. Thus for a standard deviation $\sigma$ of 10\% of $\tau$; i.e. $w=0.1$ we have upper and lower values of $A$ within approximately ±0.02 of their respective base values. For a standard deviation $\sigma$ of 20\% of $\tau$; i.e. $w=0.2$ we have upper and lower values of $A$ within approximately ±0.04 of their respective base values. This confirms that the findings arrived at analytically for the periodic arrival of new information are relevant even in realistic scenarios.
We now use these models in an Agent Based Modeling framework. We perform two simulations. The first simulation for $N=500$ and $\tau = 5$ hrs corresponds to $P=0.072 < P_{2}$ i.e. all agents expected to be on the surface with fold, i.e. polarized. The second simulation for $N = 100$ and $\tau = 100$ hrs corresponds to $P=36 > P_{1}$ i.e., all agents expected to be fully on the surface without fold, i.e. no polarization. We initialize $I$ \& $O$ of agents randomly and uniformly and let several iterations run. Figure \ref{fig:image_last} plots opinion (O) vs information (I) for all $N$ agents at the beginning and after several iterations. Figure \ref{fig:image_last}(b) for $N = 500$ and $\tau = 5$ hrs shows that agents are polarized, with no agent having near neutral opinion. Figure \ref{fig:image_last}(d) for $N = 100$ and $\tau = 100$ hrs shows agents have opinions spread over the entire range from $-1$ to $+1$, i.e. no polarization. This confirms that the models when used in an Agents Based Modeling framework produce the same outcome as was derived analytically. We have used the same $\tau$ for all agents within each simulation only because it is simple to set up.
\subsection*{5. Conclusions}
Opinion formation in individuals is cast as a dynamical system. The dynamical system is used to study the effect of social media that has made it possible for large groups of individuals to stay connected and communicate faster, on polarization. Polarization number and extent of polarization are defined. Polarization number is a function of sociological parameters, viz, time interval at which new information is received by an individual and the size of the individual's network. Critical values of the Polarization number are in turn functions of psychological parameters. We show that staying connected with large groups of people and frequent consumption of information may have a role in making individuals prone to polarization and may have played a role in polarizing society.
This finding is significant. Critical values of the polarization parameter arrived at here correspond to values of the psychological parameter in \cite{van_der_mass}. The findings provide a framework to fine-tune these parameter values to align with the evidence. If fine-tuning of parameters is not adequate to explain the evidence then this provides a framework to explore changes in model structure by psychology and sociology experts. Results can also help individuals to introspect and purposefully move their respective polarization numbers to reduce the propensity for polarization.
Actual details of the society post-polarization, like how many individuals are in each polarized group and how far removed the two groups are, etc can be obtained using agent-based modeling.
\section*{Acknowledgements}
Authors are thankful for the encouragement from Cisco in the form of payment of Article Process Charges associated with publication of this paper. We are thankful to many who have made access to literature easy and free. We are thankful to Han L J van der Maas, Jonas Dalege \& Lourens Waldorp for a highly readable paper that motivated us to formulate and attempt this study. We look forward to the opportunity to collaborate with experts from core disciplines of Opinion Dynamics. The first author (Sudhakar) is thankful to Mr M.R. Ganesh for extending a conducive and excellent environment for working on this problem.
\nolinenumbers


\begin{thebibliography}{99}
\bibitem{gaitonde}
J. Gaitonde, J. Kleinberg, and É. Tardos, 
``Polarization in Geometric Opinion Dynamics,'' 
in \textit{Proceedings of the 22nd ACM Conference on Economics and Computation (EC '21)}, 
Budapest, Hungary, 2021, pp. 499--519.
\bibitem{musco2021quantify}
C. Musco, I. Ramesh, J. Ugander, and R. T. Witter, 
``How to Quantify Polarization in Models of Opinion Dynamics,'' 
arXiv preprint arXiv:2110.11981, 2021.
\bibitem{Noorazar_2020}
Noorazar, H., Vixie, K., Talebanpour, A. \& Hu, Y. From classical to modern opinion dynamics. {\em International Journal Of Modern Physics C}. \textbf{31}, 2050101 (2020,7), https://doi.org/10.1142
\bibitem{opinion_dynamics_review2}
Xia, H., Wang, H. \& Xuan, Z. Opinion Dynamics: A Multidisciplinary Review and Perspective on Future Research. {\em IJKSS}. \textbf{2} pp. 72-91 (2011,10)
\bibitem{lorenz2007continuous}
Lorenz, J. Continuous opinion dynamics under bounded confidence: A survey. {\em International Journal Of Modern Physics C}. \textbf{18}, 1819-1838 (2007)
\bibitem{urena2019review}
Urena, R., Kou, G., Dong, Y., Chiclana, F. \& Herrera-Viedma, E. A review on trust propagation and opinion dynamics in social networks and group decision making frameworks. {\em Information Sciences}. \textbf{478} pp. 461-475 (2019)
\bibitem{anderson2019recent}
Anderson, B. \& Ye, M. Recent advances in the modelling and analysis of opinion dynamics on influence networks. {\em International Journal Of Automation And Computing}. \textbf{16}, 129-149 (2019)
\bibitem{leonard2021nonlinear}
Leonard, N., Lipsitz, J., Pitt, R. \& Bertschinger, N. Nonlinear dynamics of the public opinion. {\em Frontiers In Physics}. \textbf{8} (2021)
\bibitem{lorenz-spreen2021}
Lorenz-Spreen, P., Lewandowsky, S., Sunstein, C. \& Hertwig, R. How behavioural sciences can promote truth, autonomy and democratic discourse online. {\em Nature Human Behaviour}. \textbf{4}, 1102-1109 (2020)
\bibitem{biondi2023dynamics}
Biondi, B., Baldovin, F. \& Stella, A. Dynamics of radicalization under different social network topologies. {\em Scientific Reports}. \textbf{13}, 1479 (2023)
\bibitem{lee2016}
Lee, F. Hong Kong citizens’ trust in China’s central government: assessing the roles of the news media, political cynicism and value dissimilarity. {\em China Quarterly}. \textbf{228} pp. 1070-1089 (2016)
\bibitem{allcott2020}
Allcott, H., Braghieri, L., Eichmeyer, S. \& Gentzkow, M. The welfare effects of social media. {\em American Economic Review}. \textbf{110}, 629-676 (2020)
\bibitem{vanbavel2021}
Van Bavel, J., Baicker, K., Boggio, P., Capraro, V., Cichocka, A., Cikara, M., Crockett, M., Crum, A., Douglas, K., Druckman, J., Drury, J., Dube, O., Ellemers, N., Finkel, E., Fowler, J., Gelfand, M., Han, S., Haslam, S., Jetten, J., Kitayama, S., Mobbs, D., Napper, L., Pennycook, G., Peters, E., Petty, R., Rand, D., Reicher, S., Schnall, S., Shariff, A., Skitka, L., Smith, S., Sunstein, C., Tabri, N., Tucker, J., Linden, S., Lange, P., Weeden, K., Wohl, M., Zaki, J., Zion, S. \& Willer, R. Using social and behavioural science to support COVID-19 pandemic response. {\em Nature Human Behaviour}. \textbf{4}, 460-471 (2020)
\bibitem{strogatz1994nonlinear}
Strogatz, S. Nonlinear dynamics and chaos: with applications to physics, biology, chemistry, and engineering. (CRC Press,1994)
\bibitem{scheffer2022belief}
Scheffer, M. \& Van Der Maas, H. Belief in models: on the fragility of convex relationships. {\em Evolutionary Human Sciences}. \textbf{4} (2022)
\bibitem{van_der_mass}
van der Maas, H., Kolstein, R., Jansen, B. \& Kievit, R. The double learning process in belief dynamics. {\em PLoS One}. \textbf{15}, e0242230 (2020)
\bibitem{Abe_2017}
Abe, S. On the Poincare-Bendixson theorem. {\em Journal Of Physics: Conference Series}. \textbf{936}, 012001 (2017)
\bibitem{stanley1987introduction}
Stanley, H. Introduction to phase transitions and critical phenomena. (Oxford University Press,1987)
\bibitem{BALL20021}
Ball, P. The physical modelling of human social systems. {\em Complexus}. \textbf{1}, 190-206 (2002,12)
\bibitem{grasman2010fitting}
Grasman, R., van der Maas, H. \& Wagenmakers, E. Fitting the cusp catastrophe in R: A cusp package primer. {\em Journal Of Statistical Software}. \textbf{32} (2010)
\bibitem{van2003sudden}
Van Der Maas, H. \& Molenaar, P. A sudden transition in the acquisition of a response selection task. {\em Journal Of Experimental Psychology: Learning, Memory, And Cognition}. \textbf{19}, 1255 (2003)
\bibitem{isermann2011identification}
Isermann, R. Identification of dynamical systems. (Springer,2011)
\bibitem{daffin2021principles}
Daffin, L. Principles of social psychology. (Pressbooks,2021)
\bibitem{james1890principles}
James, W. The principles of psychology. (Holt,1890)
\bibitem{zajonc1968attitudinal}
Zajonc, R. Attitudinal effects of mere exposure. {\em Journal Of Personality And Social Psychology}. \textbf{9}, 1-27 (1968)
\end{thebibliography}
\end{document}